# Suppression of bend loss in writing of three-dimensional optical waveguides with femtosecond laser pulses

ZHENGMING LIU[1,2,3], YANG LIAO[1,7], ZHIWEI FANG[1,2,3], WEI CHU[1] AND YA CHENG[1,4,5,6,8]

[1]*State Key Laboratory of High Field Laser Physics, Shanghai Institute of Optics and Fine Mechanics, Chinese Academy of Sciences, Shanghai 201800, China*
[2]*University of Chinese Academy of Sciences, Beijing 100049, China*
[3]*School of Physical Science and Technology, Shanghai Tech University, Shanghai 200031, China*
[4]*State Key Laboratory of Precision Spectroscopy, East Normal University, Shanghai 200062, China*
[5]*Collaborative Innovation Center of Extreme Optics, Shanxi University, Taiyuan, Shanxi 030006, China*
[6]*XXL – The Extreme Optoelectromechanics Laboratory, East China Normal University, Shanghai 200062, China*
[7]*superliao@vip.sina.com*
[8]*ya.cheng@siom.ac.cn*

**Abstract:** We provide a solution toward compact and low-loss three dimensional (3D) photonic circuits by femtosecond laser direct writing of 3D waveguides in fused silica. We suppress the bend loss by sandwiching the waveguide between a pair of walls formed by internal modification of glass. Our method allows to reduce the bend loss of a curved waveguide with a bending radius of 15 mm by more than one order of magnitude.

## 1. Introduction

Optical waveguides are one of the fundamental building blocks for integrated photonic applications [1]. Recently, femtosecond laser has been used for inscribing optical waveguides in various transparent materials such as glass, crystals and polymers thanks to the unique capability of locally modifying refractive index in the vicinity of focal spot [2-6]. This has allowed producing three-dimensional (3D) photonic circuits for a broad spectrum of applications ranging from quantum information processing and miniaturized lasers to optomechanics and optofluidics [7-10]. Typically, a refractive index change on the order of $\sim 10^{-4}$-$\sim 10^{-3}$ can be achieved by irradiating the interior of glass or crystal with tightly focused femtosecond laser pulses, which is sufficient for producing single-mode waveguides of low propagation losses. Increasing the refractive index change requires stronger modification in the laser irradiated regions, which can be realized with either higher laser fluence or optimization of composition of the substrate materials. Unfortunately, due to the extreme nonlinear nature of femtosecond laser interaction with transparent materials, the modification realized under the strong irradiation condition can always lead to high propagation losses. For instance, fluctuation in the laser peak power and randomly distributed defects in the material will cause severe inhomogeneity in the regions irradiated with intense laser pulses, particularly when the laser power is significantly higher than the threshold power required for initiating the photoionization. The difficulty in inducing large refractive index changes smoothly distributed in the laser irradiated regions is a major obstacle for producing compact photonic devices because of the limitation imposed by high bend losses at small radii of curvature [11].

Although femtosecond laser written waveguides with high refractive index contrast have been demonstrated in several materials, such as nanocrystal-doped glass [12], phase-separable and leachable glass [13], borate laser crystals [14], alkaline earth boro-aluminosilicate glass [15], and PMMA polymers [16], fused silica is an ideal substrate material for optical and photonic applications. Particularly, waveguides inscribed in fused silica can support propagation losses as low as 0.1 dB/cm at 1550 nm wavelength [17]. Nevertheless, waveguide-based photonic circuits produced by femtosecond laser direct writing in fused silica still suffer from the issue of large bend losses at small radii of curvatures. Here, we solve this challenging problem by creating a peculiar sandwich structure at the bend of waveguides. By sandwiching the curved waveguide between a pair of vertical walls produced within glass by femtosecond laser irradiation, the refractive index change in the guiding area can be substantially increased. The walls will be called as bend-loss-suppression walls (BLSWs) in the following. We also optimize the geometrical parameters of the BLSWs for minimizing the bend loss.

## 2. Experimental

In this work, a Ti:Sapphire regenerative amplifier (Libra-HE, Coherent Inc.) with an operation wavelength of 800 nm, a pulse width of ~50 fs, and a repetition rate of 1 kHz was used. The laser was focused into polished fused silica glass (Corning 7980) using an objective lens with a numerical aperture (N.A.) of 0.8 (MPlanFLN 50×, Olympus). The waveguide was fabricated with transverse writing scheme, i.e., the scan direction was perpendicular to the laser propagation direction. To obtain a symmetrical cross section of the waveguide, the femtosecond laser beam was shaped with an adaptive slit mapped onto the pupil plane of the objective lens by a phase-only spatial light modulator (SLM, Hamamatsu, X10468-02). For writing low-loss sing-mode curved waveguides, orientation of the adaptive slit was varied during the writing so as to remain parallel to the sample translation direction [18]. For both the waveguides and BLSWs, a circularly polarized femtosecond laser beam was used to circumvent the influence of nanogratings induced by linearly polarized femtosecond laser [19], and the laser writing was always performed along a same direction to avoid the directional quill effect [20].

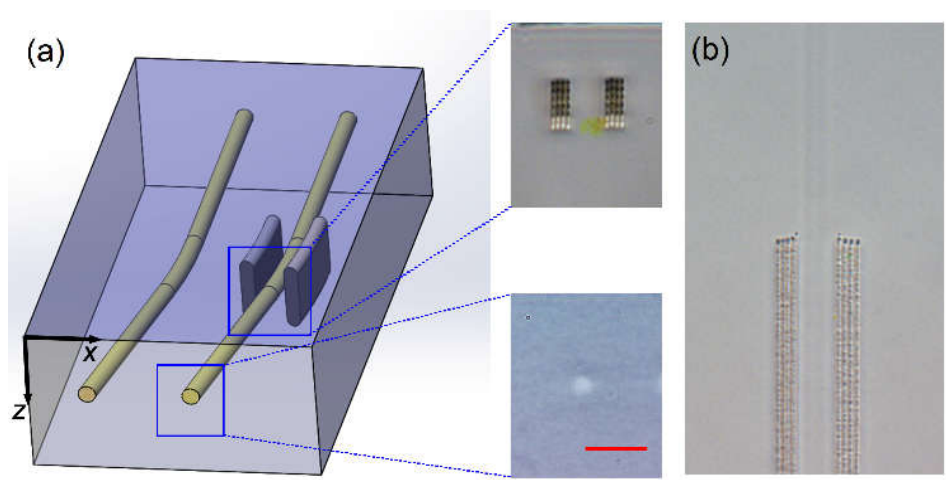

Fig. 1. (a) The 3D configuration of the waveguide and modification structures, the insets show the cross sections of the waveguide and BLSW. Scale bar: 30 μm. (b) A top view micrograph of a section of curved waveguide sandwiched by a pair of BLSWs.

The design of the 3D waveguides is illustrated in Fig. 1. The waveguide inscribed 50 μm below surface is of a length of ~2 cm, which consists of two straight sections connected by a bend section. The arc angle of bend was 5° and the bend radius was 15 mm. The whole waveguide, no matter straight or curved, was written with a single scan of the slit-shaped beam at a fixed scan speed of 0.03 mm/s. The width and length of the slit were respectively set at 0.44 mm and 2 mm, and the pulse energy measured after the slit aperture was ~1.9 μJ.

The BLSW consists of a number of parallel modification tracks, which was written with a circular beam of 5-mm diameter at a pulse energy of 0.3 μJ and a writing speed of 0.03 mm/s. A matrix writing method was used to control the cross-section of BLSW. Along the propagation direction (the Z direction in Fig.1), the offset between adjacent tracks was set at ~4.5 μm, and the number of tracks was 5. In the horizontal direction (the X direction in Fig.1), the offset between adjacent tracks was set at ~3 μm, and the number of layers varied from 1 to 5.

After inscription of both the optical waveguides and BLSWs, the end facets of samples were grounded and polished. The waveguide mode profile and insertion loss were characterized with a 780 nm laser source, fed by a single mode polarization maintaining fiber (PMF) for end-fire coupling into the waveguide. The loss of bend section was obtained from the difference between transmitted powers of the waveguides with and without the bend section. The near–field mode profile at the exit facet was imaged onto a CCD camera with an objective lens of

NA = 0.45. By measuring the mode-field diameter (MFD) in the far field at different distances from the end facet of the waveguides, the NA of waveguide can be determined. To estimate the loss along the waveguide sections, the scattered light along the waveguide carrying the He-Ne laser (633 nm) was also recorded with a top-view CCD camera. The scattered intensity along the straight waveguide was used to determine the propagation loss with an exponential fit.

## 3. Results and discussions

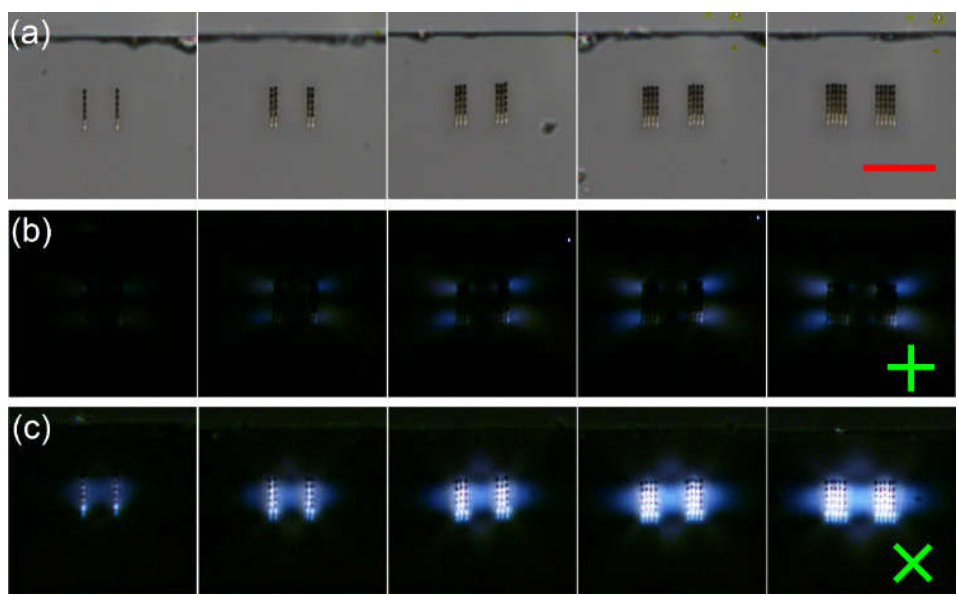

Fig. 2. Bright field (a) and polarized light microscopy (b, c) images of the BLSWs with different layers. The separation between two inner walls is fixed to be 22 μm (center to center), and the layer number of modification tracks ranges from 1 to 5 from left to right. (b) Images with two crossed polarizers parallel and perpendicular to the glass surface. (c) Images with two crossed polarizers both on 45° angle with the glass surface. The green crosses in (b, c) indicate the orientation of the polarizers. All the micrographs were taken under the same illumination condition. Scale bar: 50 μm.

In fused silica, the stress-induced refractive index change by a single modification track is typically on the order of ~$10^{-5}$-~$10^{-4}$ [21-22]. To produce stronger stress distribution, multiple modification tracks were arranged into an array to form the modification wall. Figure 2 show the stress distributions induced by the BLSWs of different numbers of layers. The results were obtained using transmitted polarized light microscope. Note that all the crossed polarizing micrographs were taken under the same illumination condition, enabling qualitative comparison of the strengths of stress. The stress was significantly enhanced by the multilayer BLSWs. In particular, the stress in the horizontal direction between the separated walls shown in Fig.2(c) predominates over the original stress generated in individual waveguide as shown in Fig. 2(b), which indicates an increase of the refractive index in this region [23]. Similar effect was also found in PMMA polymer [16].

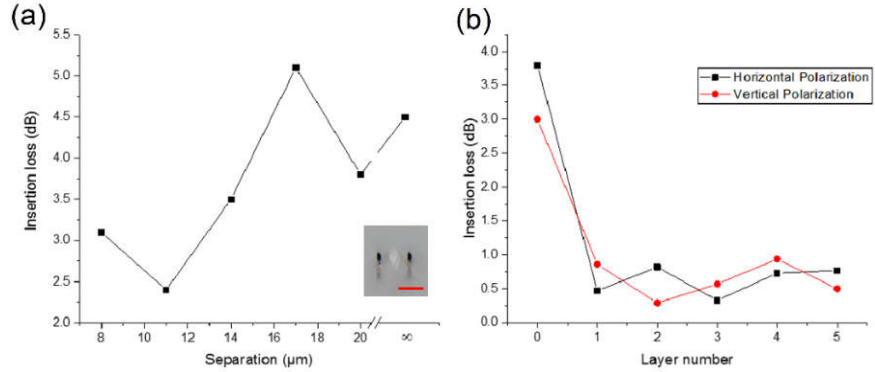

Fig. 3. (a) The influence of the separation between the waveguide and the modification tracks (center to center) on the bend loss of curved waveguides, and the black square at ∞ separation represents the reference bend loss measured for a curved waveguide without BLSWs. The inset shows the corresponding modification region. Scale bar: 20 μm; (b) The influence of the layer number on the loss of curved waveguides. The values at a layer number of 0 represent the reference losses measured for a curved waveguide without BLSWs.

To minimize the bend loss, we examine the influence of several geometrical parameters of BLSW on the loss of curved waveguides. We begin by adding only a pair of single-scan modification tracks along the curved waveguides and investigate how the bend loss depends on the distance of the separated walls. Figure 3(a) shows that the insertion loss of curved waveguides is quite sensitive to the separation distance between the two tracks. A minimal loss of 2.5 dB has been achieved when the modification tracks on the two sides of the waveguide are equally separated from the center of waveguide by a distance of ~ 11 μm. It should be stressed that although the stress-induced refractive index change increases with the decreasing separation, the scattering loss also becomes high at small distances between the modification tracks and waveguide due to micro-roughness in the modification tracks. Therefore a tradeoff has to be made in choosing the separation distance between the modification tracks.

Notably, the dependence of the bend loss on the number of layers appears to be irregular. This could be understood as a result of the non-negligible evanescent coupling between the waveguides and the BLSWs at variable separation distance. Due to this, the separation of ~ 17 μm results in a higher bend loss even than that of the curved waveguide alone. Similar phenomena were previously reported by Fernandes et al., in which the modification tracks were used to tune the birefringence of waveguides [21].

Figure 3(b) shows that the influence of the number of layers of BLSWs on the bend loss of curved waveguides. According to the result in Fig. 3(a), the separation distance between BLSW and waveguide is fixed at 11 μm (center to center). It can be seen that the bend losses of curved waveguides decreases dramatically by adding the BLSWs. The minimal insertion losses around ~0.3 dB were achieved with two-layer and three-layer BLSWs for both the vertical and horizontal polarization modes.

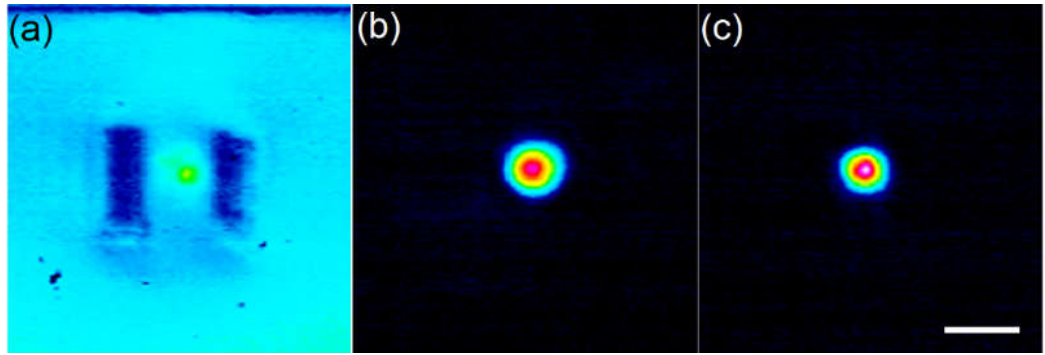

Fig. 4. (a) Mode profile under bright field showing the position of guiding mode between the BLSWs; and mode profiles at 780 nm wavelength for straight waveguide without (b) and with (c) four-layer BLSWs. Scale bar: 20 μm.

For comparison, the mode profiles at 780 nm from the straight waveguides without and with the BLSWs are shown in Fig. 4(b) and 4(c), respectively. The waveguide mode (MFD = 15.0 μm×16.2 μm) becomes smaller and more symmetric (MFD=12.1 μm×12.6 μm) after adding the BLSWs. The refractive-index changes in the straight waveguides without and with the BLSWs were estimated by measuring the numerical aperture (NA) of the cone of light emerging from the end facet of the waveguides. For small refractive index change (Δn) and assuming a step index waveguide, the refractive index change could be calculated by $\Delta n = NA^2/2n$. The results show the refractive index change was increased from $1.85 \times 10^{-3}$ to $3.45 \times 10^{-3}$ after adding a pair of four-layer BLSWs on both sides of the waveguide.

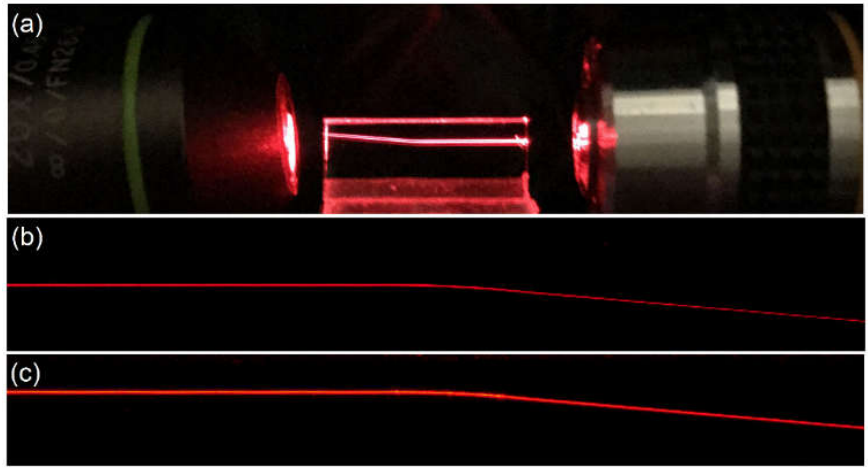

Fig. 5. (a) The photograph of waveguide carrying 633nm beam. Top view images of a waveguide bend without (b) and with (c) BLSWs. The total length of waveguide is ~2 cm and the curvature radius of waveguide bend is 15 mm.

The bend loss of curved waveguides includes the radiation loss, the propagation loss, and the mode mismatch losses at the transitions between the straight and bend segments. Although the BLSWs introduces higher scattering loss and mode mismatch losses, the radiation loss drops dramatically following an exponentially decay with the increasing refractive index change (Δn) [24]. To observe directly the loss along the bend section, the scattered light along the waveguide carrying the He-Ne laser (633 nm) was recorded with a top-view CCD camera, as shown in Fig. 5. Clearly, light propagation suffers lower loss by adding the BLSWs. The exponential decay of the scattering light along the waveguide indicates a propagation loss of ~0.6 dB/cm at 633 nm in the straight section of the waveguide. The total propagation loss of the waveguide of a length of ~2 cm and a sharp bend of 15 mm was estimated to be ~ 1.5 dB, including the 1.2-dB loss in the straight section and 0.3-dB loss at the bend.

## 4. Conclusions

In conclusion, we have demonstrated suppression of bend loss of 3D waveguides written in fused silica by femtosecond laser direct writing. This is achieved by sandwiching the bend segment between a pair of BLSWs. By optimizing the location and geometrical parameters of the BLSWs, the bend loss of curved waveguides with a bending radius of 15 mm was reduced from 3~4 dB to ~0.3 dB. Our technique will benefit high-density integration of waveguides into compact photonic circuits.

**Acknowledgments**

This work is supported by the National Natural Science Foundation of China (Grant Nos. 61590934, 61675220, 1173409, 11674340, 61327902), National Basic Research Program of China (Grant No. 2014CB921303), the Strategic Priority Research Program of Chinese Academy of Sciences (Grant No. XDB16000000), the Key Research Program of Frontier Sciences, Chinese Academy of Sciences (Grant No. QYZDJ-SSW-SLH010), and the Project of Shanghai Committee of Science and Technology (Grant 17JC1400400).